\documentclass[aps,prd,preprint,preprintnumbers,unsortedaddress,superscriptaddress,nofootinbib,longbibliography]{revtex4-1}
\usepackage{subfigure}
\usepackage{graphicx}
\usepackage{epsfig}
\usepackage{amsmath}
\usepackage{amsfonts}
\usepackage{amssymb}
\usepackage{xcolor}
\usepackage{slashed}
\usepackage{lipsum}
\usepackage{comment}
\usepackage{epstopdf}
\usepackage{hyperref}
\usepackage{float}
\usepackage{wrapfig}
\usepackage[utf8]{inputenc}
\usepackage{siunitx} 
\usepackage{comment}
\usepackage{soul}
\usepackage{cancel}
\usepackage{mathtools}
\mathtoolsset{showonlyrefs=true}
\hypersetup{colorlinks = true,
           allcolors = blue}

\newcommand{\Slash}[1]{{\ooalign{\hfil/\hfil\crcr$#1$}}}
\newcommand{\tento}[1]{\times 10^{#1}}

\newcommand{\cc}{{\rm c.c.}}

\newcommand{\im}{{\rm Im}}
\newcommand{\re}{{\rm Re}}
\newcommand{\br}{{\rm Br}}

\newcommand{\kev}{{\rm keV}}
\newcommand{\mev}{{\rm MeV}}
\newcommand{\gev}{{\rm GeV}}

\newcommand{\mL}{\mathcal{L}}
\newcommand{\mM}{\mathcal{M}}

\newcommand{\mN}{\mathcal{N}}
\newcommand{\itp}{\affiliation{CAS Key Laboratory of Theoretical Physics, Institute of Theoretical Physics,\\ Chinese Academy of Sciences,  Beijing 100190, China}}
\newcommand{\ucas}{\affiliation{School of Physical Sciences, University of Chinese Academy of Sciences,\\ Beijing 100049, China}}
\graphicspath{{figures/}}
\allowdisplaybreaks[1]
\newcommand{\ee}{{e^+e^-}}
\newcommand{\ppb}{{p\bar{p}}}
\newcommand{\X}{X(3872)}
\newcommand{\Z}{Z_c(4020)}
\newcommand{\dtri}{D_\triangle}
\begin{document}
\title{Possible precise measurements of the \texorpdfstring{$X(3872)$}{x} mass with the \texorpdfstring{$e^+e^-\to\pi^0\gamma X(3872)$}{eetopi0gamx} and \texorpdfstring{$p\bar p\to\gamma X(3872)$}{ppbtogamx} reactions}
\date{\today}
\author{Shuntaro Sakai}
\email{shsakai@mail.itp.ac.cn}
\itp
\author{Hao-Jie Jing}
\email{jinghaojie@itp.ac.cn}
\itp
\ucas
\author{Feng-Kun Guo}
\email{fkguo@itp.ac.cn}
\itp
\ucas 

\begin{abstract}
It was recently proposed that the $X(3872)$ binding energy, the difference between the $D^0\bar D^{*0}$ threshold and the $X(3872)$ mass, can be precisely determined by measuring the $\gamma X(3872)$ line shape from a short-distance $D^{*0}\bar D^{*0}$ source produced at  high-energy experiments. Here, we investigate the feasibility of such a proposal by estimating the cross sections for the $e^+e^-\to\pi^0\gamma X(3872)$ and $p\bar p\to\gamma X(3872)$ processes considering the $D^{*0}\bar D^{*0}D^0/\bar D^{*0}D^{*0}\bar D^0$ triangle loops.
These loops can produce a triangle singularity  slightly above the $D^{*0}\bar D^{*0}$ threshold.
It is found that the peak structures originating from the  $D^{*0}\bar D^{*0}$ threshold cusp and the triangle singularity are not altered much by the energy dependence introduced by the $e^+e^-\to\pi^0D^{*0}\bar D^{*0}$ and $p\bar p\to\bar D^{*0}D^{*0}$ production parts or by considering a finite width for the $X(3872)$.
We find that $\sigma(e^+e^-\to\pi^0\gamma X(3872)) \times \br(X(3872)\to\pi^+\pi^-J/\psi)$ is $\mathcal{O}(0.1~{\rm fb})$ with the $\gamma X(3872)$ invariant mass integrated from 4.01 to 4.02~GeV and the c.m. energy of the $e^+e^-$ pair fixed at 4.23~GeV.
The cross section $\sigma(p\bar p\to\gamma X(3872))\times \br(X(3872)\to\pi^+\pi^-J/\psi)$ is estimated to be of $\mathcal{O}(10~{\rm pb})$. 
Our results suggest that a precise measurement of the $X(3872)$ binding energy can be done at PANDA.
\end{abstract}

\maketitle

\section{Introduction}
\label{sec:intro}

Among many charmoniumlike states listed in the Review of Particle Physics (RPP)~\cite{Zyla:2020zbs}, 
special attention has been paid to the $\X$.\footnote{In this paper, the $\chi_{c1}(3872)$ in the RPP~\cite{Zyla:2020zbs} is denoted by $X(3872)$ or merely $X$, and $\Z$ or $Z_c$ stands for the $X(4020)$ in the RPP.}
The mass of $\X$ is consistent with the $D^0\bar D^{*0}$ threshold energy, $m_X=(3871.69\pm 0.17)~\mev$, and only an upper bound is provided for its small width, $\Gamma_X<1.2~\mev$~\cite{Zyla:2020zbs}. The latest experimental development comes from the LHCb collaboration that reported precise determinations of the mass and width~\cite{Aaij:2020qga,Aaij:2020xjx}. In particular, a detailed analysis of the $X(3872)$ line shape using the  Flatt\'e parametrization~\cite{Flatte:1976xu}, which is more proper than the Breit-Wigner (BW) form for states near an $S$-wave strongly coupled threshold, is performed in Ref.~\cite{Aaij:2020qga}.
The closeness of its mass and the $D^0\bar D^{*0}$ threshold invokes the hadronic molecular description of $\X$:
the $\X$ is treated as a shallow $S$-wave bound state of $D\bar D^*$, e.g., in Refs.~\cite{Tornqvist:1993ng,Braaten:2004fk,Gamermann:2007fi,Fleming:2007rp,AlFiky:2005jd,Ding:2009vj,Dong:2009yp,Li:2012cs,Guo:2013zbw}.
Such a description can successfully explain the large branching ratio of the isospin forbidden $\X\to\pi^+\pi^-J/\psi$ relative to the isospin allowed $\pi^+\pi^-\pi^0J/\psi$ mode~\cite{Gamermann:2009fv},
and the strong coupling of the molecular state to its constituents in the molecular description, i.e., $\X$ to $D\bar D^*$,
would naturally explain the large branching fractions of the $\X$ to $\pi^0D^0\bar D^0/D^0\bar D^{*0}$~\cite{Zyla:2020zbs,Li:2019kpj,Braaten:2019ags}.
Many works are devoted to elucidate the composition of the $\X$ from its decay properties~\cite{Braaten:2003he,Swanson:2003tb,Swanson:2004pp,Braaten:2005ai,Dong:2008gb,Dong:2009uf,Nielsen:2010ij,Mehen:2011ds,Guo:2014taa}.
The strong coupling of the $X(3872)$ to the $D^0\bar D^{*0}$ in an $S$-wave implies that there must be a strong cusp exactly at the threshold~\cite{Guo:2019twa}, complicating the line shape analysis. 
The line shapes of the $\pi^+\pi^-J/\psi$ and/or $D^0\bar D^{*0}$ distributions were analyzed with the Flatt\'e parametrization~\cite{Hanhart:2007yq,Braaten:2007dw,Zhang:2009bv,Kalashnikova:2009gt,Aaij:2020qga} or the effective range expansion~\cite{Stapleton:2009ey,Kang:2016jxw} in which the threshold effect is incorporated by requiring unitarity; however, no conclusive results for the nature of the $\X$ have been achieved so far.
See, e.g., Refs.~\cite{Esposito:2014rxa,Lebed:2016hpi,Guo:2017jvc,Kalashnikova:2018vkv,Yamaguchi:2019vea,Brambilla:2019esw} and references therein for further information on works related to $\X$.

Recently, a possible way to precisely determine the $\X$ binding energy, which is defined as the difference between the $D^0\bar D^{*0}$ threshold and the $X(3872)$ mass\footnote{A negative $\delta$ corresponds to a mass above the threshold and thus a resonant state in this paper.}
\begin{align}
    \delta=m_{D^0}+m_{\bar D^{*0}}-m_X,\label{eq:delx}
\end{align} 
was proposed in Ref.~\cite{Guo:2019qcn}. This can be done by measuring the $\gamma\X$ distribution instead of the $\X$ line shape in its decay products like $\pi^+\pi^-J/\psi$ or $D^0\bar D^{*0}$.
Consider a triangle diagram for the transition of an $S$-wave $D^{*0}\bar D^{*0}$ pair, produced at short distances in some high-energy experiment, into $\gamma\X$. The $D^{*0}$ ($\bar D^{*0}$) subsequently decays into $\gamma D^0$ ($\gamma \bar D^0$), and the $\X$ is produced by merging the $D^0\bar D^{*0}+\bar D^0 D^{*0}$ pair at the last step.
The process thus proceeds via a  $D^{*0}\bar D^{*0}D^0$ triangle loop.
This loop can have a triangle singularity (TS) due to the simultaneous on-shellness of all three intermediate mesons, which leads to a peak in the $\gamma\X$ distribution just above the $D^{*0}\bar D^{*0}$ threshold.
With the Landau equation~\cite{Landau:1959fi} or with a simple equation for the TS position derived with a refined formulation~\cite{Bayar:2016ftu},
one sees that the TS position is sensitive to the $\X$ mass: 
the TS is located at $4015.14~\mev$ with $\delta=-180~\kev$ and  $4015.64~\mev$ with $\delta=-50~\kev$.
For the $\X$ mass within $(3871.69\pm0.17)$~MeV~\cite{Zyla:2020zbs}, the TS appears in the range of $m_{\gamma X}\in[4015.17,4016.40]~\mev$ which can be obtained by using Eqs.~(55) and (60) in Ref.~\cite{Guo:2019twa}.
While the TS, at which the amplitude diverges logarithmically, is turned into a finite peak due to the width of the internal particles,
the peak originating from the TS of the $D^*\bar D^*D$ loop should be still clear thanks to the tiny width of the $D^{*0}$, which is only $55.3\pm1.4~\kev$~\cite{Rosner:2013sha,Guo:2019qcn}.
Then, one expects that the $\X$ binding energy can be determined well with the precise measurement of the TS peak in the $\gamma\X$ distribution.

The role of the TS stemming from the $D^{*}\bar D^{*}D$ loop on the $\X$ production has been studied in some papers.
The $\ee\to\gamma\X$ transition 
is studied in Refs.~\cite{Braaten:2019gwc,Braaten:2020iye}, and the $Y(4260)\to\X\gamma$ decay is studied in Ref.~\cite{Dong:2014zka} by including the contribution of $J/\psi\rho$, $J/\psi\omega$, and the compact component made of $c\bar c$ explicitly.
In Ref.~\cite{Voloshin:2019ivc}, the energy dependence of the $Z_c(4020)^0\to\gamma\X$ branching fraction is studied.
One can see the difference of the energy dependence by changing the $\X$ binding energy.
In addition to the radiative reactions, decays emitting a pion with the $D^{*0}\bar D^{*0}D^0$ loop have also been considered~\cite{Voloshin:2019ivc,Braaten:2019yua,Sakai:2020ucu}.
While the TS appears in a smaller range of the $\pi\X$ energy compared with the $\gamma\X$ case,
the asymmetry of the $\pi X(3872)$ line shape may be used to extract the $X(3872)$ binding energy.
The decay process $B\to(J/\psi\pi^+\pi^-)K\pi$ with the $J/\psi\pi^+\pi^-$ produced by the $D^0\bar D^{*0}$ rescattering considering the $D^{*+}\bar D^{*0}D^0/D^{*-}D^{*0}\bar D^0$ loop is studied in Ref.~\cite{Nakamura:2019nwd}.
For more works related to the TS, we refer to Ref.~\cite{Guo:2019twa}.

In this paper, we investigate two promising reactions in which the proposal of precisely measuring the $\X$ binding energy by virtue of the TS mechanism may be realized:
the $\ee\to\pi^0\gamma\X$ and $\ppb\to\gamma\X$ reactions.
In these reactions, the $D^*\bar D^*$ pair can be produced in an $S$ wave.
In the case of the $\ee$ collisions, the isovector resonance $\Z$ seen in the $D^*\bar D^*$ distribution of the $\ee\to\pi^0(D^*\bar D^*)^0$ process~\cite{Ablikim:2015vvn}
is expected to be a good source of the $S$-wave $D^*\bar D^*$ pair,
and high-statistics data can be expected for the $\ppb$ reaction by the PANDA experiment at the Facility for Antiproton and Ion Research (FAIR) in the near future.

This paper is organized as follows.
In Sec.~\ref{sec:formalism}, the formalism for calculating the $\ee\to\pi^0\gamma\X$ and $\ppb\to\gamma\X$ amplitudes is provided where the effect of the $\X$ width is taken into account.
The results of our calculation, the $\gamma\X$ invariant mass distributions in these reactions and the estimated cross sections, are given in Sec.~\ref{sec:results}.
A brief summary is given in Sec.~\ref{sec:summary}.
Detailed expressions of the amplitudes used in Sec.~\ref{sec:formalism} are relegated to Appendix~\ref{app:1}.

\section{Formalism}
\label{sec:formalism}
\subsection{\texorpdfstring{$\ee\to\pi^0\gamma\X$}{eetopi0gamx}}
\label{sec:eetopgx}

\begin{figure}[t]
 \centering
 \includegraphics[width=9cm]{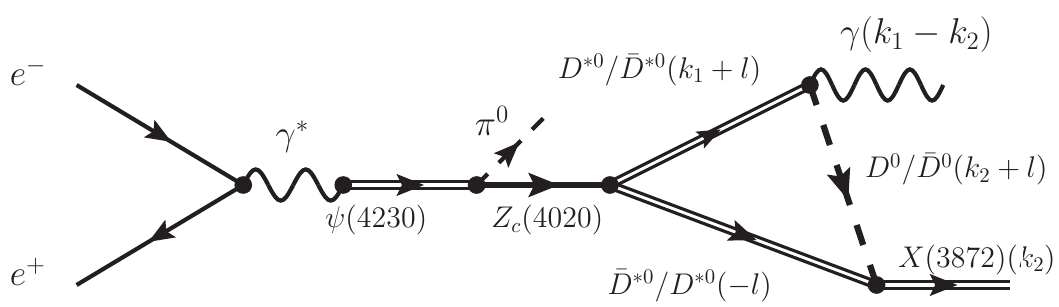}
 \caption{Triangle diagram contributing to the $\ee\to\pi^0\gamma\X$ process considered here.}
 \label{fig:fig1}
\end{figure}

First, we consider the $\ee\to\pi^0\gamma\X$ amplitude with the $D^{*0}\bar D^{*0}D^0/\bar D^{*0}D^{*0}\bar D^0$ loops.
The diagram is given in Fig.~\ref{fig:fig1}.
Only the neutral $D^*\bar D^*D/\bar D^*D^*\bar D$ loops are accounted for the process
because we focus on the TS peak of the $\gamma\X$ invariant mass distribution near the $D^{*0}\bar D^{*0}$ threshold
and the $\X$ appears near the $D^0\bar D^{*0}$ threshold as a narrow peak.
As found in Ref.~\cite{Ablikim:2015vvn},
the $(D^*\bar D^*)^0$ distribution of $\ee\to\pi^0(D^*\bar D^*)^0$ at the c.m. energies $\sqrt{s}=4.23$ and 4.26~GeV can be described well by including a resonance with $J^{P}=1^+$,
and the $(D^*\bar D^*)^0$ pair is predominantly produced by the resonance around the $D^*\bar D^*$ threshold.
Here, we also assume that the $\Z$ is the $J^P=1^+$ exotic state which can decay into an $S$-wave $D^*\bar D^*$ pair.
The $\pi\Z$ pair is produced by the $\psi(4230)$ resonance, which is seen in some hidden- and open-charm productions~\cite{Zyla:2020zbs} and would be needed to describe the dependence of the cross section on the $\ee$ c.m. energy 
because the $\ee\to\pi D^{*}\bar D^{*}$ cross section at $\sqrt{s}=4.26~\gev$ is smaller than that of $\sqrt{s}=4.23~\gev$~\cite{Ablikim:2015vvn}.
We use the central values of the mass and width of the $\psi(4230)$ given in the RPP~\cite{Zyla:2020zbs}, $m_{\psi}=(4220\pm 15)~\mev$ and $\Gamma_\psi=(60\pm 40)~\mev$.
Note that, while the width of the $\psi(4230)$ is not fixed well, the $\gamma\X$ invariant mass distribution at a given $\sqrt{s}$, which will be considered in this work, is not affected by the details of the $\psi(4230)$ properties.

The $\ee\to\gamma^*$, $\gamma^*\to\psi(4230)$, $\psi(4230)\to\pi^0\Z^0$, and $\Z^0\to D^{*}\bar D^{*}$ amplitudes are written as follows:
\begin{align}
 -it_{\ee,\gamma}=&\, ieg_{\mu\nu}\bar{v}\gamma^\mu u(\epsilon_\gamma^*)^\nu,\\
 -it_{\gamma,\psi}=&\, ieg_0g_{\mu\nu}(\epsilon_\gamma)^\mu(\epsilon_{\psi}^*)^\nu, \\
 -it_{\psi,\pi^0Z_c}=&\, ig_1g_{\mu\nu}(\epsilon_\psi)^\mu(\epsilon_{Z_c}^*)^\nu, \\
 -it_{Z_c,D^{*}\bar D^{*}}=&\, ig_2\epsilon^{\mu\nu\rho\sigma}(p_{Z_c})_\mu(\epsilon_{Z_c})_\nu(\epsilon_{D^{*}}^*)_\rho(\epsilon_{\bar D^{*}}^*)_\sigma,
\end{align}
where $e(e>0)$ denotes the electric charge unit, $\bar v$ and $u$ are the spinors for the positron and electron, respectively, and the $\epsilon$'s are the polarization vectors of the involved spin-1 particles.
With the isospin symmetry and the phase convention $\left|D^{(*)+}\right>=-\left|I=1/2,I_z=1/2\right>$,
a minus sign is needed for the $\Z^0\to D^{*+}D^{*-}$ coupling constant relative to the $\Z^0\to D^{*0}\bar D^{*0}$ coupling.
Constant amplitudes are used for the $S$-wave vertices of the $\psi(4230)\to \pi^0\Z^0$ and $\Z^0\to D^{*}\bar D^*$ because the lowest angular momentum gives the dominant contribution in the near-threshold region.
Then, the $\ee\to\pi^0D^{*}\bar D^{*}$ amplitude is given by
\begin{align}
 -i\mM_{\ee,\pi^0D^{*}\bar D^{*}}=&\, ie^2g_0g_1g_2D_\gamma^{-1}(s)D_\psi^{-1}(s)D_{Z_c}^{-1}(m_{D^{*}\bar D^{*}}^2)\notag\\
 &\times\bar{v}\gamma_{\beta''}u[P_\psi]^{\beta''\beta'}[P_{Z_c}]_{\beta'\beta}\epsilon^{\alpha\beta\gamma\delta}(p_{Z_c})_\alpha(\epsilon_{D^{*}}^*)_\gamma(\epsilon_{\bar D^{*}}^*)_\delta\\
 \equiv&-i\mM_{\ee,\pi^0D^{*}\bar D^{*}}^{\gamma\delta}(\epsilon_{D^{*}}^*)_\gamma(\epsilon_{\bar D^{*}}^*)_\delta,\label{eq:1}
\end{align}
with $D_R(s)=s-m_R^2+im_R\Gamma_R$ and $[P_R]^{\mu\nu}=-g^{\mu\nu}+\frac{p_R^\mu p_R^\nu}{m_R^2}$.
The energy dependence of the width is taken into account as done in Ref.~\cite{Ablikim:2015vvn} (see also the review on the resonances of Ref.~\cite{Zyla:2020zbs}):
\begin{align}
 \Gamma_{Z_c}(m_{D^*\bar D^*})=&\, \frac{\Gamma_{Z_c0}}{2}
 \left(\frac{p_{D^{*0}}(m_{D^*\bar D^*})}{p_{D^{*0}}(m_{Z_c0})}+\frac{p_{D^{*+}}(m_{D^*\bar D^*})}{p_{D^{*+}}(m_{Z_c0})}\right),\\
 p_{D^{*}}(m_{D^*\bar D^*})=&\, \frac{1}{2m_{D^*\bar D^*}}\lambda^{1/2}(m_{D^*\bar D^*}^2,m_{D^{*}}^2,m_{\bar D^{*}}^2),
\end{align}
where $\lambda(x,y,z)=x^2+y^2+z^2-2xy-2yz-2zx$.
The central values of $m_{Z_c0}$ and $\Gamma_{Z_c0}$ in Ref.~\cite{Ablikim:2015vvn}, $m_{Z_c0}=(4031.7\pm 2.1)~\mev$ and $\Gamma_{Z_c0}=(25.9\pm 8.8)~\mev$, are used.
With the amplitude in Eq.~\eqref{eq:1}, the differential cross section of $\ee\to\pi^0Z_c(4020)^0\to\pi^0(D^{*}\bar D^{*})^0$, $d\sigma_{\ee,\pi^0(D^*\bar D^*)^0}/dm_{D^*\bar D^*}$, is given by
\begin{align}
 \frac{d\sigma_{\ee,\pi^0(D^*\bar D^*)^0}}{dm_{D^*\bar D^*}}=&\, 
 \sum_{D^*\bar D^*}\frac{p_{\pi^0}p_{D^*}}{(4\pi)^5p_es}\int d\Omega_{\pi^0}\int d\Omega_{D^{*}}\overline{|\mM_{\ee,\pi^0D^*\bar D^*}|^2},\label{eq:dcs0}
\end{align}
with $p_{\pi^0}=\lambda^{1/2}(s,m_{\pi^0}^2,m_{D^*\bar D^*}^2)/(2\sqrt{s})$, $p_{D^*}=\lambda^{1/2}(m_{D^*\bar D^*}^2,m_{D^*}^2,m_{\bar D^*}^2)/(2m_{D^*\bar D^*})$, and $p_e=\lambda^{1/2}(\sqrt{s},m_e^2,m_e^2)/(2\sqrt{s})$.
The sum of $D^*\bar D^*$ takes care of both the $D^{*0}\bar D^{*0}$ and $D^{*+}D^{*-}$ that are included in the $(D^*\bar D^*)^0$ final state observed by BESIII~\cite{Ablikim:2015vvn}.
The solid angles $\Omega_{\pi^0}$ and $\Omega_{D^*}$ are those in the $\ee$ c.m. frame and $D^*\bar D^*$ c.m. frame, respectively.
The overlined quantities are those after the spin sum and average.
With the $\ee\to\pi^0Z_c(4020)^0\to\pi^0(D^*\bar D^*)^0$ cross section in Ref.~\cite{Ablikim:2015vvn}, $(61.6\pm 8.2)$~pb at $\sqrt{s}=4.23~\gev$,
the product of the coupling constant $g_0g_1g_2$ is fixed to be $g_0g_1g_2=0.68~\gev^3$.

Now, we move to the $D^{*0}\bar D^{*0}D^0$ triangle loop amplitude.
The $P$-wave $D^{*0}\to\gamma D^0$ transition amplitude is given by~\cite{Casalbuoni:1996pg}
\begin{align}
 -i\mM_{D^{*0},\gamma D^0}= eg_3\epsilon^{\mu\nu\rho\sigma}(p_{D^{*0}})_\mu(p_\gamma)_\nu(\epsilon_{D^{*0}})_\rho(\epsilon_\gamma^*)_\sigma,\label{eq:dstdgam}
\end{align}
and the parameter $g_3$ is fixed to be $g_3=1.77~\gev^{-1}$ with the $D^{*0}\to\gamma D^0$ branching ratio $35.3\%$~\cite{Zyla:2020zbs} and the $D^{*0}$ full width $\Gamma_{D^{*0}}=55.3~\kev$~\cite{Guo:2019qcn},
which can be obtained by using isospin symmetry to relate to the $D^{*+}$ full width and the $D^{*+}\to\pi^+D^0$ and $D^{*0}\to\pi^0D^0$ branching ratios~\cite{Braaten:2015tga,Guo:2019qcn}.\footnote{The coupling constant of $D^{*+}\to\gamma D^{+}$, $g_3'$, evaluated with the measured full width and branching ratio is $g_3'=0.47~\gev^{-1}$, which is less than 1/3 of the $D^{*0}\to\gamma D^0$ coupling. This makes the charged $D^{*}\bar D^*D$-loop contribution even less important.}
The $\bar D^{*0}\to \gamma \bar D^0$ amplitude needs one minus sign that comes from the $C$ parity of the photon and the convention of the $C$ transformation, $CD^{*0}=+\bar D^{*0}$.

\begin{figure}[t]
 \centering
 \includegraphics[width=8cm]{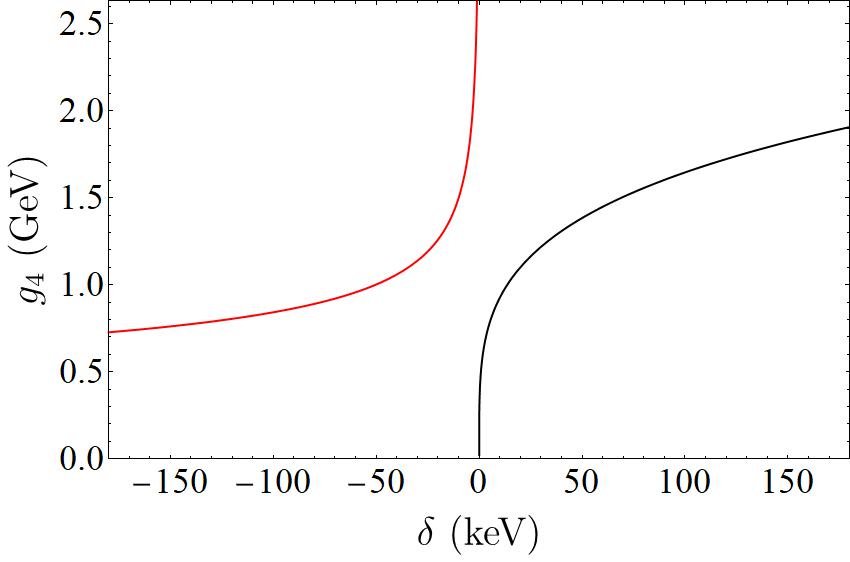}
 \caption{The coupling constant $g_4$ as a function of the $\X$ binding energy, $\delta$.
 The black and red lines in $\delta>0$ and $\delta<0$ correspond to the cases with the $\X$ mass below and above the $D^0\bar D^{*0}$ threshold,
 and $g_4$ is evaluated with Eqs.~\eqref{eq:gx1} and \eqref{eq:gx2}, respectively.}
 \label{fig:coup1}
\end{figure}

The $S$-wave transition amplitude of the $D^0\bar D^{*0}\to\X$ transition is written as
\begin{align}
 -it_{D^0\bar D^{*0},X}=ig_4g_{\mu\nu}(\epsilon_{\bar D^{*0}})^\mu(\epsilon_X^*)^\nu,\label{eq:ddstx}
\end{align} 
and the coupling constant of $\bar D^0D^{*0}\to\X$ is the same.
We estimate the coupling constant $g_4$ with two different ways for the $\X$ mass above or below the $D^{0}\bar D^{*0}$ threshold.
When the $\X$ mass is below the $D^0\bar D^{*0}$ threshold, the coupling constant can be evaluated assuming the $\X$ is an $S$-wave $D^0\bar D^{*0}$ molecule~\cite{Baru:2003qq,Gamermann:2009uq,Lin:2017mtz},
\begin{align}
 g_X^2=\frac{16\pi m_X^2}{\mu_{D^0\bar D^{*0}}}\sqrt{2\mu_{D^0\bar D^{*0}}\delta},~\label{eq:gx1}
\end{align}
with $\mu_{D^0\bar D^{*0}}$ and $\delta$ being the $D^0\bar D^{*0}$ reduced mass and the $\X$ binding energy given by Eq.~\eqref{eq:delx}, respectively.
In Eq.~\eqref{eq:gx1}, $g_X$ is the coupling constant of $\X$ to the $D\bar D^*$ pair of $J^{PC}=1^{++}$, and $g_4$ and $g_X$ are related with $g_4=g_X/2$~\cite{Sakai:2020ucu}.
With the analyses of the $\X$ line shape in the $\pi^+\pi^-J/\psi$ or $D^0\bar D^{*0}$, the non-$D^0\bar D^{*0}$ component of $X(3872)$ is estimated to be a few tens of percents~\cite{Braaten:2005ai,Kalashnikova:2009gt,Aaij:2020qga}, which would give uncertainties of the same level to the $X(3872)\to D\bar D^*$ coupling squared evaluated with Eq.~\eqref{eq:gx1}.
When the $\X$ mass is above the $D^0\bar D^{*0}$ threshold, $g_4$ can be obtained by using the $\X\to D^0\bar D^{*0}$ branching ratio~\cite{Lees:2019xea};
using Eq.~\eqref{eq:ddstx}, we have
\begin{align}
 g_4^2=\frac{1}{2}\Gamma_{X}\br[\X\to D^{*0}\bar D^0+\cc]\frac{8\pi m_X^2/p_{D^0}}{\frac{2}{3}\left(1+\frac{E_{\bar D^{*0}}^2}{2m_{\bar D^{*0}}^2}\right)}\label{eq:gx2}
\end{align}
with $p_{D^0}=\lambda^{1/2}(m_X^2,m_{D^0}^2,m_{D^{*0}}^2)/(2m_X)$ and $E_{\bar D^{*0}}=(m_X^2+m_{\bar D^{*0}}^2-m_{D^0}^2)/(2m_X)$.
In this work, the mass of $\X$ is treated as a parameter, and it will be changed to see the difference of the $\gamma\X$ invariant mass distribution.
The width of $\X$, $\Gamma_X$, is currently not known and the upper bound is provided~\cite{Zyla:2020zbs}.
Here, $\Gamma_X$ is assumed to be $100~\kev$, which is the value expected from the calculation of the $\X\to\pi^0D^0\bar D^0$ partial width in the hadronic molecular picture~\cite{Fleming:2007rp,Guo:2014hqa,Dai:2019hrf} and the $\X\to\pi^0D^0\bar D^0$ branching ratio~\cite{Zyla:2020zbs,Li:2019kpj,Braaten:2019ags}.\footnote{In the recent LHCb analyses~\cite{Aaij:2020qga,Aaij:2020xjx}, the $\X$ BW parameters are extracted from the $\pi^+\pi^-J/\psi$ distribution: the mass is consistent with the $D^0\bar D^{*0}$ threshold within the error and the width is about $1~\mev$. If $\Gamma_X=1~\mev$, the nontrivial cuspy structure in the $\gamma\X$ distribution would get largely smeared, and it would be difficult to obtain clear information of the $\X$ from the $\gamma\X$ line shape. However, it should be emphasized that the BW form is not suitable for such a state seated on top of an $S$-wave threshold as the energy dependence of the width and its analytic continuation, which is crucial under this situation, is not properly taken into account. In the best fit with the Flatt\'e parametrization, the half-maximum width of the original $\X$ line shape before taking account of the experimental resolution is of the order of $100~\kev$~\cite{Aaij:2020qga}, which is compatible with the molecular model prediction~\cite{Fleming:2007rp,Guo:2014hqa,Dai:2019hrf}.}
The coupling constant $g_4$ as a function of $\delta$ is shown in Fig.~\ref{fig:coup1}.
Note that the values of $g_4$ from both the $\delta>0$ and $\delta<0$ sides are similar if we neglect the part with $\delta$ in the vicinity of $0$. In that special region, the absolute value of the imaginary of the pole position cannot be approximated by half the width computed using Eq.~\eqref{eq:gx2}.
Furthermore, the coupling of the $\X$ to the charged and neutral $D\bar D^*$ can be computed from the residue of the coupled-channel $D^0\bar D^{*0}$--$D^+D^{*-}$ $T$-matrix. It is found that the couplings of the $\X$ to $D^0\bar D^{*0}$ and to $D^+D^{*-}$ are approximately the same~\cite{Gamermann:2009fv,Guo:2014hqa}, and are consistent with the values shown in Fig.~\ref{fig:coup1} (see also the discussion in Ref.~\cite{Sakai:2020ucu}). 
{In the end, we use $g_4=1~\gev$ for all the cases of the $\X$ masses that will be discussed below for an estimation of the cross section.}

Then, with the amplitudes Eqs.~\eqref{eq:1}, \eqref{eq:dstdgam}, and \eqref{eq:ddstx}, the $\ee\to\pi^0\gamma\X$ production amplitude considering the $D^{*0}\bar D^{*0}D^0$ and $\bar D^{*0}D^{*0}\bar D^0$ triangle loops in Fig.~\ref{fig:fig1} is given by
\begin{align}
 -i\mM_{\ee,\pi^0\gamma X}=&2\int\frac{d^4l}{(2\pi)^4}(-i\mM_{\ee,\pi^0D^{*0}\bar D^{*0}}^{\gamma\delta})eg_3g_4\dtri^{-1}[P_{D^{*0}}]_{\gamma\rho}[P_{\bar D^{*0}}]_{\delta\tau}\notag\\
 &\times\epsilon^{\mu\nu\rho\sigma}(p_{D^{*0}})_\mu(p_\gamma)_\nu(\epsilon_\gamma^*)_\sigma(\epsilon_X^*)^\tau\label{eq:ampeepgx},\\
 \dtri=&[l^2-m_{\bar D^{*0}}^2+i\epsilon][(k_1+l)^2-m_{D^{*0}}^2+i\epsilon][(k_2+l)^2-m_{D^0}^2+i\epsilon].
\end{align}
The factor of $2$ in the above equation comes from the same contribution from the charge-conjugated loops.
The library \textsc{LoopTools} is used for the evaluation of the one-loop integral~\cite{Hahn:1998yk}.
The width of the particles is taken into account by replacing the mass of $D^{*0}$ and $\bar D^{*0}$, $m_{D^{*0}}$, with $m_{D^{*0}}-i\Gamma_{D^{*0}}/2$ in the propagator.
See Appendix~\ref{app:1} for the details of $\mM_{\ee,\pi^0\gamma X}$.

With the $\ee\to\pi^0\gamma\X$ amplitude in Eq.~\eqref{eq:ampeepgx}, the $\gamma\X$ invariant mass distribution is given by
\begin{align}
 \frac{d\sigma_{\ee,\pi^0\gamma X}}{dm_{\gamma X}}=\frac{p_{\pi^0}p_\gamma}{(4\pi)^5sp_e}\int d\Omega_{\pi^0}\int d\Omega_{\gamma}\overline{|\mM_{\ee,\pi^0\gamma X}|^2},\label{eq:dsigee}
\end{align}
where $p_{\pi^0}$ and $p_e$ are given by the expressions below Eq.~\eqref{eq:dcs0} changing $m_{D^*\bar D^*}^2$ to $m_{\gamma X}^2$, and $p_\gamma=\lambda^{1/2}(m_{\gamma X}^2,0,m_X^2)/(2m_{\gamma X})$.
$\Omega_{\pi^0}$ and $\Omega_{\gamma}$ are the solid angles of the $\pi^0$ in the $\ee$ c.m. frame and of the photon in the $\gamma\X$ c.m. frame, respectively.

\subsection{\texorpdfstring{$\ppb\to\gamma\X$}{ppbtogamx}}

\begin{figure}[t]
 \centering
 \includegraphics[width=6cm]{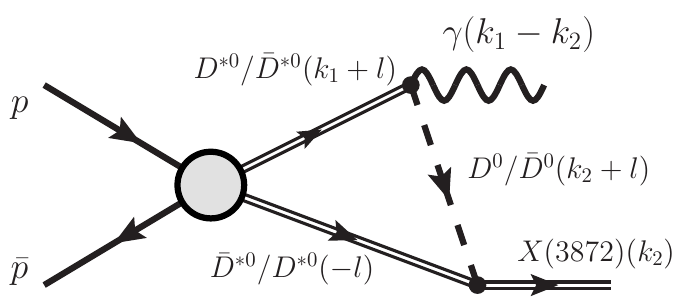}
 \caption{Triangle diagram contributing the $\ppb\to\gamma\X$ process.}
 \label{fig:ppbtogamX}
\end{figure}

The $\ppb\to\gamma\X$ amplitude is considered in this part.
The diagram of the $\ppb\to\gamma\X$ transition with the $D^{*0}\bar D^{*0}D^0/\bar D^{*0}D^{*0}\bar D^0$ loops is shown in Fig.~\ref{fig:ppbtogamX}.

\begin{figure}[t]
 \centering
 \includegraphics[width=5cm]{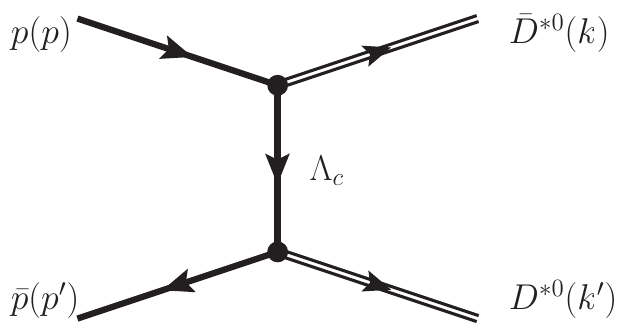}
 \caption{$\bar D^*D^*$ production from $\ppb$ through a $\Lambda_c$ exchange.
 The momenta of the particles are given in parentheses.}
 \label{fig:ppbdstdstb}
\end{figure}

The $\bar D^{*0}D^{*0}$ pair can be produced from $\ppb$ by exchanging a $\Lambda_c$ as depicted in Fig.~\ref{fig:ppbdstdstb}.
Possible $\Sigma_c^{(*)}$ contributions are ignored as argued in Ref.~\cite{Haidenbauer:2014rva} based on the flavor SU(4) model.
With the effective Lagrangian for the $p\Lambda_cD^{*0}$ coupling~\cite{Dong:2014ksa},
\begin{align}
 \mL_{N\Lambda_cD^{*0}}=g_v\bar{\Lambda}_c\gamma^\mu(D^{*0})_\mu p+{\rm h.c.},
\end{align}
the $\ppb\to \bar D^{*0}D^{*0}$ transition amplitude with the $\Lambda_c$ exchange is written as
\begin{align}
 -i\mM_{\ppb,\bar D^{*0}D^{*0}}=&\bar{v}(ig_v\gamma^{\mu'})\frac{iF_{p,\bar D^*\Lambda_c}^2}{\Slash{p}-\Slash{k}-m_{\Lambda_c}+i\epsilon}(ig_v\gamma^{\mu})u(\epsilon_{D^{*0}}^*)_{\mu'}(\epsilon_{\bar D^{*0}}^*)_{\mu},\label{eq:ppbdstdstb1}
\end{align}
where 
$u$ and $\bar{v}$ are the spinors of the proton and antiproton, and a form factor $F_{p,\bar D^*\Lambda_c}$ is introduced.
For the parameter $g_v$, we take the value in Refs.~\cite{Liu:2001ce,Dong:2014ksa} obtained by using the SU(4) model, $g_{v}=-5.20$.
For the form factor $F_{p,\bar D^*\Lambda_c}$, we use
\begin{align}
 F_{p,\bar D^*\Lambda_c}^2=\frac{\Lambda^4}{((p-k)^2-m_{\Lambda_c}^2)^2+\Lambda^4}.\label{eq:ff}
\end{align}
The form factor like Eq.~\eqref{eq:ff} is used, e.g., in Refs.~\cite{He:2015yva,Lin:2017mtz},
and the cutoff is typically set to be around $\Lambda=2~\gev$.
Here, since the aim is to get an order-of-magnitude estimate of the cross section for the $p\bar p\to \gamma \X$, it suffices to take a value used in the literature, and we take $\Lambda=2.0~\gev$. 
The dependence of our results on this parameter will be checked.

We are interested in the manifestation of the TS in the $\gamma\X$ invariant mass distribution.
As shown in Ref.~\cite{Coleman:1965xm}, the TS emerges when the process can occur classically, 
i.e., the internal particles of the loop are simultaneously placed on shell and all the momenta are collinear.
At this time, the exchanged $\Lambda_c$ in the $\bar D^{*0}D^{*0}$ production is far away from on shell.
Then, Eq.~\eqref{eq:ppbdstdstb1} can be approximated by taking the leading term of the expansion in powers of $1/m_{\Lambda_c}$. The $\ppb\to\bar D^{*0}D^{*0}$ production amplitude is reduced to
\begin{align}
 -i\mM_{\ppb,\bar D^{*0}D^{*0}}=&\frac{ig_v^2F_{p,\bar D^*\Lambda_c}^2}{m_{\Lambda_c}}
 \bar{v}\gamma^{\mu'}\gamma^{\mu}u(\epsilon_{D^{*0}}^*)_{\mu'}(\epsilon_{\bar D^{*0}}^*)_{\mu}\notag\\
 \equiv&-i\mM_{\ppb,\bar D^{*0}D^{*0}}^{\mu'\mu}(\epsilon_{D^{*0}}^*)_{\mu'}(\epsilon_{\bar D^{*0}}^*)_{\mu}.\label{eq:ppbdstdstb2}
\end{align}
Because the internal particles are close to on shell in the vicinity of the TS energies,
the 4-momentum transfer $(p-k)^2$ in $F_{p,\bar D^*\Lambda_c}^2$ can be approximated by
\begin{align}
 (p-k)^2=m_p^2+m_{\bar D^{*0}}^2-2m_{\bar D^{*0}}E_p,
\end{align}
where the spatial momentum of the $\bar D^{*0}$ is ignored because the TS energy is close to the $D^{*0}\bar D^{*0}$ threshold.

The part of the triangle loop in Fig.~\ref{fig:ppbtogamX} is the same as the $\ee\to\pi^0\gamma\X$ reaction given in Sec.~\ref{sec:eetopgx}.
The $\ppb\to\gamma\X$ amplitude with the $D^{*0}\bar D^{*0}D^0$ loop is written as
\begin{align}
 -i\mM_{\ppb,\gamma X}^{(D^{*0}\bar D^{*0}D^0)}
 =&\int\frac{d^4l}{(2\pi)^4}(-i\mM_{\ppb,\bar D^{*0}D^{*0}}^{\mu'\mu})eg_3g_4\dtri^{-1}[P_{D^{*0}}]_{\mu'\gamma}[P_{\bar D^{*0}}]_{\mu\tau}\notag\\
 &\epsilon^{\alpha\beta\gamma\delta}(p_{D^{*0}})_\alpha(p_\gamma)_\beta(\epsilon_\gamma^*)_\delta(\epsilon_X^*)^\tau,\label{eq:ppbt1}
\end{align}
and the $\bar D^{*0}D^{*0}\bar D^0$ loop gives
\begin{align}
 -i\mM_{\ppb,\gamma X}^{(\bar D^{*0}D^{*0}\bar D^0)}
 =&\int\frac{d^4l}{(2\pi)^4}(-i\mM_{\ppb,\bar D^{*0}D^{*0}}^{\mu'\mu})(-eg_3g_4)\dtri^{-1}[P_{\bar D^{*0}}]_{\mu\gamma}[P_{D^{*0}}]_{\mu'\tau}\notag\\
 &\epsilon^{\alpha\beta\gamma\delta}(p_{\bar D^{*0}})_\alpha(p_\gamma)_\beta(\epsilon_\gamma^*)_\delta(\epsilon_X^*)^\tau.\label{eq:ppbt2}
\end{align}
The details of $\mM_{\ppb,\gamma X}^{(D^{*0}\bar D^{*0}D^0/\bar D^{*0}D^{*0}\bar D^0)}$ can be found in Appendix~\ref{app:1}.
Finally, the amplitude of the $\ppb\to\gamma\X$ with the $D^{*0}\bar D^{*0}D^0/\bar D^{*0}D^{*0}\bar D^0$ loops, $\mM_{\ppb,\gamma X}$, is 
\begin{align}
 \mM_{\ppb,\gamma X}=\mM_{{\rm ISI}}(\mM_{\ppb,\gamma X}^{(D^{*0}\bar D^{*0}D^0)}+\mM_{\ppb,\gamma X}^{(\bar D^{*0}D^{*0}\bar D^0)}),~\label{eq:ppbamp1}
\end{align}
where $\mM_{{\rm ISI}}$ is a factor to take into account the $\ppb$ initial-state interaction (ISI).
In Ref.~\cite{Dong:2014ksa}, this factor $|\mM_{\rm ISI}|^2$ is about $0.25$ at $\sqrt{s}=5~\gev$ and moderately increases along with $\sqrt{s}$.
Here we treat $\mM_{\rm ISI}$ as a constant and take $|\mM_{\rm ISI}|^2=0.2$ for an estimation of the ISI effect.

With the $\ppb\to\gamma\X$ amplitude given in Eq.~\eqref{eq:ppbamp1} and the phase-space factor, 
the cross section of the $\ppb\to\gamma\X$, $\sigma_{\ppb,\gamma X}$, as a function of $\sqrt{s}$, which is now the $\ppb$ c.m. energy, is given by
\begin{align}
 \sigma_{\ppb,\gamma X}=\int d\Omega\frac{1}{64\pi^2s}\frac{k}{p}\overline{|\mM_{\ppb,\gamma X}|^2},
\end{align}
with $k=\lambda^{1/2}(s,0,m_X^2)/(2\sqrt{s})$ and $p=\lambda^{1/2}(s,m_p^2,m_p^2)/(2\sqrt{s})$.

\subsection{Width effect of the \texorpdfstring{$\X$}{x}}
\label{subsec:width}

To take into account the width of the $\X$,
the cross sections need to be convolved with the spectral function of the $\X$.\footnote{See Ref.~\cite{Ortega:2020ayw} for a detailed discussion on the smearing effect of the experimental energy resolution,
and see also Ref.~\cite{Guo:2019qcn} for arguments for the sensitivity of the TS peak on the $\X$ binding energy, where the binning of the $\gamma\X$ energy is considered.
}
The spectral function may be parametrized using either the BW or the Flatt\'e form. 
{The latter form for the spectral function is used in this work since the Flatt\'e parametrization is more proper for analyzing the $\X$ line shape which is very close to the $D^0\bar D^{*0}$ threshold. The spectral function with the Flatt\'e parametrization} is given by~\cite{Hanhart:2007yq,Aaij:2020qga}
\begin{align}
 \rho_X(\tilde{m}_X)=&-\frac{1}{\pi}\im\left(\frac{1}{D_X}\right),D_X={\tilde{m}_X-m_{X0}+i\Gamma_X(\tilde{m}_X)/2},\label{eq:flattedist}\\
 \Gamma_X(\tilde{m}_X)=&g(k_1+k_2)+\Gamma_{X,\rho}(\tilde{m}_X)+\Gamma_{X,\omega}(\tilde{m}_X)+\Gamma_{X0},\\ 
 k_1=&\sqrt{2\mu_{D^0\bar D^{*0}}(\tilde{m}_X-m_{D^0}-m_{\bar D^{*0}})}, \\
 k_2=&\sqrt{2\mu_{D^+D^{*-}}(\tilde{m}_X-m_{D^+}-m_{D^{*-}})}, \\
 \Gamma_{X,\rho}(\tilde{m}_X)=&f_\rho\int_{2m_\pi}^{\tilde{m}_X-m_{J/\psi}}\frac{dm'}{2\pi}\frac{q(\tilde{m}_X,m')\Gamma_\rho}{(m'-m_\rho)^2+\Gamma_\rho^2/4},\label{eq:gamxrho}\\
 \Gamma_{X,\omega}(\tilde{m}_X)=&f_\omega\int_{3m_\pi}^{\tilde{m}_X-m_{J/\psi}}\frac{dm'}{2\pi}\frac{q(\tilde{m}_X,m')\Gamma_\omega}{(m'-m_\omega)^2+\Gamma_\omega^2/4},\\
 q(\tilde{m}_X,m')=&\frac{1}{2\tilde{m}_X}\lambda^{1/2}(\tilde{m}_X^2,m'^2,m_{J/\psi}^2),
\end{align}
with $\Gamma_\rho$ and $\Gamma_\omega$ being the widths of the $\rho$ and $\omega$ mesons, respectively.
The nonrelativistic momenta $k_{1,2}$ are analytically continued below the threshold.
In the case with the Flatt\'e amplitude, 
the scaling property hinders a determination of all free parameters~\cite{Baru:2004xg}.
We make use of the Flatt\'e parameters, $m_{X0}$, $\Gamma_{X0}$, $g$, $f_\rho$, and $f_\omega$ from Ref.~\cite{Aaij:2020qga} which fixes $m_{X0}$ and fits the other parameters to the data{, and $g_4=1~\gev$ is used for the $\X\to D^0\bar D^{*0}$ coupling as mentioned in Sec.~\ref{sec:eetopgx} for the estimation of the order of the cross section.}

As pointed out in Ref.~\cite{Guo:2019qcn}, for determining the $\X$ binding energy from the $\gamma\X$ line shape, the $\X$ needs to be reconstructed from decay modes other than the $\pi^0D^0\bar D^0$ one;
otherwise, one has to consider the tree-level contribution of $D^{*0}\bar D^{*0}\to \pi^0D^0\bar D^0$, which has a subtle interference with the triangle diagrams and cannot be treated as a smooth background near the TS energies~\cite{Schmid:1967ojm,Anisovich:1995ab,Szczepaniak:2015hya,Debastiani:2018xoi}.
In Ref.~\cite{Braaten:2020iye}, the $\ee\to\gamma D^{*0}\bar D^0$ process is studied, and it is found that 
the $D^0\bar D^{*0}$ distribution with a fixed $\sqrt{s}$ is completely dominated by the tree-level contribution, which increases rapidly at the TS energy.

{Because of the existence of a TS, in the invariant mass distribution of the decay products of the $\X$ for a given initial energy, or in the $\gamma\X$ distribution when the invariant mass of the final state particles used to reconstruct the $\X$ (such as the $J/\psi\pi^+\pi^-$) is constrained within a small region around the $D^0\bar D^{*0}$ threshold, there should be a TS peak even without the formation of the $\X$, as pointed out in Ref.~\cite{Nakamura:2019nwd}.
Such an effect would not cause any trouble, and will be automatically included if the full amplitude is employed for the transition from the $D^0\bar D^{*0}$ to the final states (such as the $J/\psi\pi^+\pi^-$) that are used to reconstruct the $\X$ (the amplitude for the complete process will be then given by a convolution of the triangle loop and the transition amplitude).
Since the transition amplitude possesses a pole due to the existence of the $\X$, it will be dominated by the $\X$ pole in the vicinity of the $\X$ mass (e.g., within $\pm2$~MeV, see Appendix~\ref{app:jpsirho}) and thus we can approximate it by the $\X$ spectral function as treated here.
}

In this work, we consider the $\pi^+\pi^-J/\psi$ mode for reconstructing the $\X$.
Then, we make the convolution as follows:
\begin{align}
 \bar{F}(m_{\gamma X})=&\int^{m_X+2\Gamma_X}_{m_X-2\Gamma_X} d\tilde{m}_X{\frac{1}{2\pi\mN}\frac{\Gamma_{X,\rho}(\tilde{m}_X)}{|D_X(\tilde{m}_X)|^2}}F(m_{\gamma X},\tilde{m}_X)\label{eq:conv1}\\
 =&{\int^{m_X+2\Gamma_X}_{m_X-2\Gamma_X} d\tilde{m}_X\rho_X(\tilde{m}_X)\frac{\Gamma_{X,\rho}(\tilde{m}_X)}{\re[\Gamma_X(\tilde{m}_X)]}F(m_{\gamma X},\tilde{m}_X)},\\
 \mN=&\int^{m_X+2\Gamma_X}_{m_X-2\Gamma_X} d\tilde{m}_X\rho_X(\tilde{m}_X),
\end{align}
with $F$ being $d\sigma_{\ee,\pi^0\gamma X}/dm_{\gamma X}$ or $\sigma_{\ppb,\gamma X}$.
The integration range for the convolution with the Flatt\'e amplitude is chosen to be twice of the half-maximum width of the peak, which is taken to be $\pm 400~\kev$ from the $D^0\bar D^{*0}$ threshold with the LHCb best-fit parameters.
\begin{figure}[t]
 \centering
 \includegraphics[width=10cm]{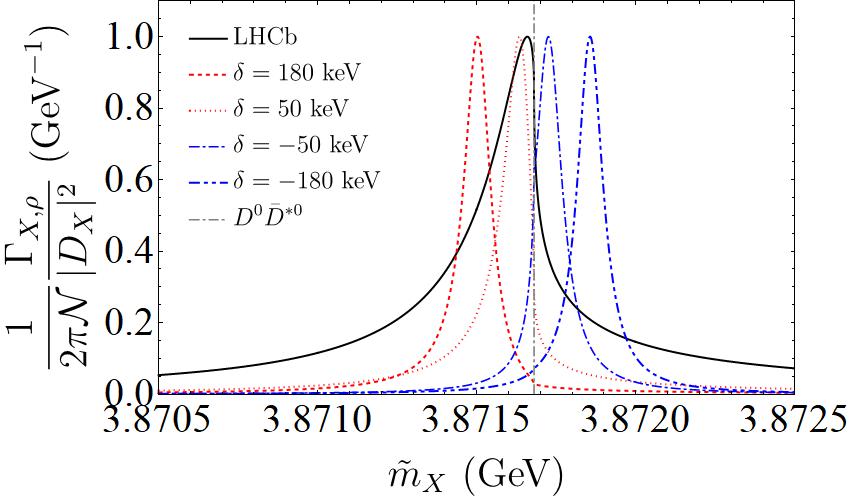}
 \caption{$(\Gamma_{X,\rho}/|D_X|^2)/(2\pi\mN)$ as functions of $\tilde{m}_X$ with
 Eq.~\eqref{eq:flattedist}.
 {The lines are normalized at their maximum values.}
 The vertical line is the $D^0\bar D^{*0}$ threshold.
 The line shape using the LHCb parameters~\cite{Aaij:2020qga} is much more asymmetric than the other choices because the corresponding pole of the Flatt\'e distribution is located in the lower half complex-$\tilde m_X$ plane but above the $D^0\bar D^{*0}$ threshold on the physical sheet of the $D^0\bar D^{*0}$ channel, defined as $\text{Im}k_1>0$ (note that the Schwarz reflection principal is not respected by the Flatt\'e distribution with a constant imaginary part, $\Gamma_{X0}$, in the denominator). For the other choices of parameters, the pole is located below threshold on the physical sheet or above threshold on the unphysical sheet, defined as $\text{Im}k_1<0$. The mass and width from all the choices are consistent with the LHCb determination within uncertainties.
 }
 \label{fig:spec}
\end{figure}
{For comparison, the calculation will also be done with different parameter sets of the Flatt\'e amplitude.
The parameters are fixed with the $\X$ binding energy and width being $\delta=\pm 180,\pm 50~\kev$ and $\Gamma_X=100~\kev$.
The mass and width are given by the peak position and the half-maximum width of $(\Gamma_{X,\rho}/|D_X|^2)/(2\pi\mN)$.
{Notice that the 100~keV width is consistent with the half-maximum width of the Flatt\'e distribution in the LHCb analysis, $0.22^{+0.26}_{-0.19}$~MeV~\cite{Aaij:2020qga}.}
The ratios of $g$, $f_\rho$, and $f_\omega$ are fixed to the same values given by the best-fit parameters with $m_{X0}=3864.5~\mev$ in Ref.~\cite{Aaij:2020qga}.
The parameters are tabulated in Table~\ref{tab:fltparameters}.}
\begin{table}[t]
 \centering
 \caption{Flatt\'e parameters of Ref.~\cite{Aaij:2020qga} with $\delta=0~\kev$ and those with $\delta=\pm 180,\pm 50~\kev$ and $\Gamma_X=100~\kev$.
 The errors of the parameters for the $\delta=0~\kev$ case given by Ref.~\cite{Aaij:2020qga} are summed in quadrature.}
 \label{tab:fltparameters}
 \begin{ruledtabular}
 \begin{tabular}[t]{c|ccccc}
 $\delta$~(keV) & $m_{X0}$~(GeV) & $g$~($-$)  & $f_{\rho}$~($-$) &  $f_\omega$~($-$) & $\Gamma_{X0}$~(MeV) \\\hline
 $0$ & $3.8645$& $0.108^{+0.006}_{-0.007}$& $(1.8^{+0.92}_{-0.85})\tento{-3}$& $1.0\tento{-2}$& $1.4\pm 0.72$ \\
 $180$ & $3.8644$& $0.097$& $1.6\tento{-3}$& $9.0\tento{-3}$& $0.0$ \\
 $50$ & $3.8643$& $0.108$& $1.8\tento{-3}$& $1.0\tento{-2}$& $0.3$\\
 $-50$ & $3.8714$& $5.186\tento{-3}$& $8.6\tento{-5}$& $4.8\tento{-4}$& $0.03$ \\
 $-180$ & $3.8717$& $2.802\tento{-3}$& $4.7\tento{-5}$& $2.6\tento{-4}$& $0.035$ \\
 \end{tabular}
 \end{ruledtabular}
\end{table}
See Fig.~\ref{fig:spec} for a plot of $(\Gamma_{X,\rho}/|D_X|^2)/(2\pi\mN)$.

Finally, the parameters used in this calculation are summarized in Table~\ref{tab:parameters}.
\begin{table}[t]
 \centering
 \caption{Parameters used in this work.}
 \label{tab:parameters}
 \begin{ruledtabular}
 \begin{tabular}[t]{ccccccc}
  $m_{D^0}$~(GeV) & $m_{D^{*0}}$~(GeV) & $\Gamma_{D^{*0}}$~(keV) & $m_{D^{*+}}$~(GeV) & $m_{\pi^0}$~(GeV) &  $m_\psi$~(GeV) & $\Gamma_\psi$~(GeV) \\
  $1.86483$ & $2.00685$ & $55.3$ & $2.01026$ & $0.13498$ & $4.22$ & $0.06$ \\\hline
  $m_p$~(GeV) &$g_0g_1g_2$~$(\gev^3)$ & $g_3$ $(\gev^{-1})$ & $m_{Z_c0}$~(GeV) & $\Gamma_{Z_c0}$~(GeV) & $g_v$ & $\Lambda$~(GeV) \\
  $0.93827$ & $0.68$ & $1.77$ & $4.0317$ & $0.0259$ & $-5.20$ & $2.0$ 
 \end{tabular}
 \end{ruledtabular}
\end{table}

\section{Results}
\label{sec:results}

\subsection{\texorpdfstring{$\ee\to\pi^0\gamma\X$}{ee to pi gamma X}}

\begin{figure}[t]
 \centering
 \includegraphics[width=8cm]{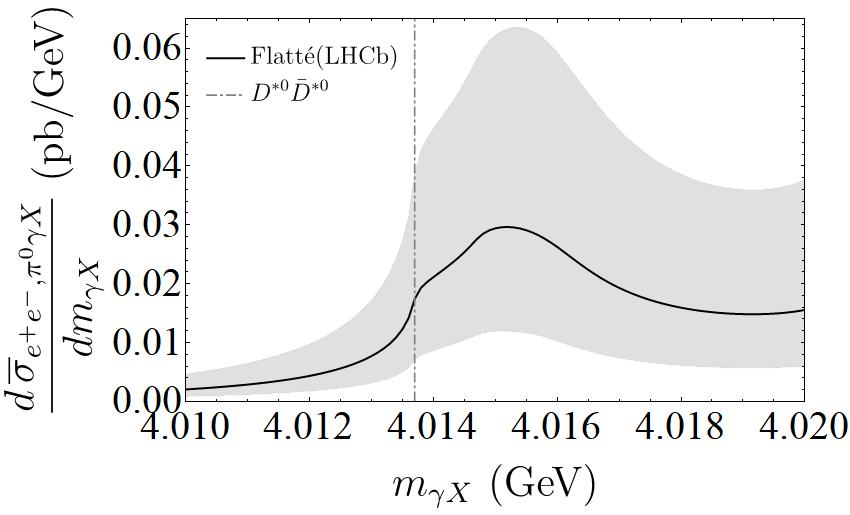}
 \includegraphics[width=8cm]{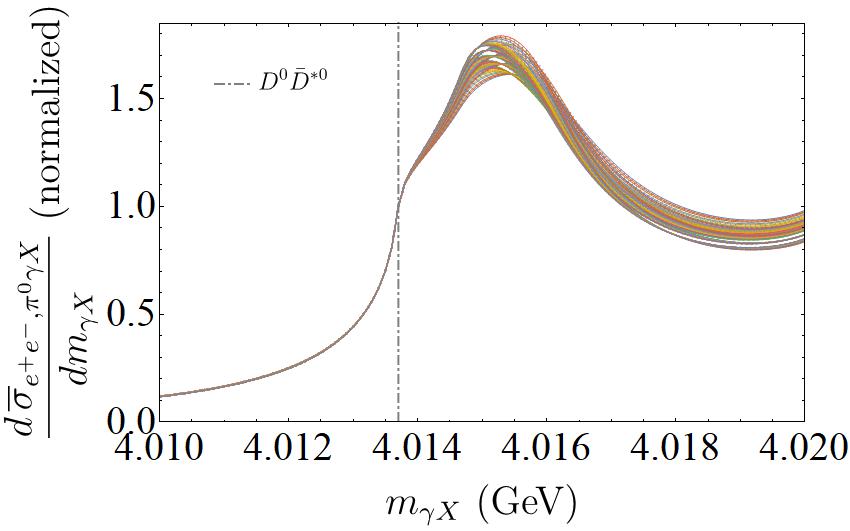}
 \caption{
 Left: the $\gamma\X$ distribution for $\ee\to\pi^0\gamma\X$ convolved with the Flatt\'e distribution, Eq.~\eqref{eq:flattedist},
 with the error band given by the parameter errors of the Flatt\'e distribution~\cite{Aaij:2020qga}.
 The $\ee$ c.m. energy is fixed at $\sqrt{s}=4.23~\gev$, and $\delta=0$~keV is used here~\cite{Aaij:2020qga}.
 {Right: the plot of the $\gamma\X$ distribution in the $\ee\to\pi^0\gamma\X$ process with the parameter sets within the Flatt\'e-parameter errors of Ref.~\cite{Aaij:2020qga} normalized with the value at the $D^{*0}\bar D^{*0}$ threshold.}
 The vertical dash-dotted line is the $D^{*0}\bar D^{*0}$ threshold in both panels.
 }
 \label{fig:res2}
\end{figure}

First, we show the $\gamma\X$ invariant mass distribution in the $\ee\to\pi^0\gamma\X$ reaction, where $X(3872)$ decays further into $\pi^+\pi^- J/\psi$, denoted by ${d\,\overline{\sigma}_{\ee,\pi^0\gamma X}/dm_{\gamma X}}$ (here and in the following, we use $\overline\sigma$ to denote cross sections convolved with the $\X$ spectral function of the $J/\psi\pi^+\pi^-$ mode).
{In order to check the impact of the uncertainty of the Flatt\'e parameters, we show the $\gamma\X$ distribution convolved with the Flatt\'e distribution Eq.~\eqref{eq:flattedist} in Fig.~\ref{fig:res2}, where $\delta$ is fixed to 0~keV.
{In the left panel of Fig.~\ref{fig:res2},} the black solid line is the result with the central values} of the best-fit parameters of the Flatt\'e analysis by LHCb in Ref.~\cite{Aaij:2020qga},
and the gray band is given by the parameter uncertainties (the statistical and systematic errors are summed in quadrature).
The peak position is 
about $4.015~\gev$, and the peak 
has a width of a few hundreds of keV.
{The cross section is averaged in the range of $\tilde m_X\in m_{D^0} + m_{D^{*0}}\pm 400~\kev$ to cover the $X(3872)$ peak region of the black solid curve in Fig.~\ref{fig:spec}.
The uncertainty from the Flatt\'e parameters is large,\footnote{We did not take into account the correlations of the parameters, and thus the error band shown here would be overestimated.} leading to a sizable uncertainty in the magnitude as seen from the gray band in Fig.~\ref{fig:res2},
but the peak position and line shape remain almost intact.
{That can be seen in the right panel of Fig.~\ref{fig:res2} for the plot with the parameter sets allowed within the errors of the Flatt\'e parameters in Ref.~\cite{Aaij:2020qga} normalized with the value at the $D^{*0}\bar D^{*0}$ threshold (see $\delta=0~\kev$ of Table~\ref{tab:fltparameters} for the value and the parameter errors); the line shapes with different parameter sets are similar to each other.}
}

Other than the TS peak, one can see a cusp of the $D^{*0}\bar D^{*0}$ threshold slightly below $m_{\gamma X}=4.014~\gev$ as a consequence of the $S$-wave production of $D^{*0}\bar D^{*0}$.
The two relevant singularities, the cusp at the $D^{*0}\bar D^{*0}$ threshold and the peak caused by the TS, fix the line shape.
The distribution shows slightly increasing behavior along with increasing $m_{\gamma X}$.
This is because of the $\Z$ resonance included in the $D^{*}\bar D^{*}$ production mechanism. Yet, its inclusion does not change the TS peak structures in the $\gamma\X$ distribution.

Notice that for the $\ee\to\gamma\X$ cross section~\cite{Braaten:2019gwc},
there is no $D^{*0}\bar D^{*0}$ threshold cusp as the $D^{*0}\bar D^{*0}$ pair is produced in $P$ wave in that case,
and only the TS peak can be seen in the $\gamma\X$ distribution. 

\begin{figure}[t]
 \centering
 \includegraphics[width=8cm]{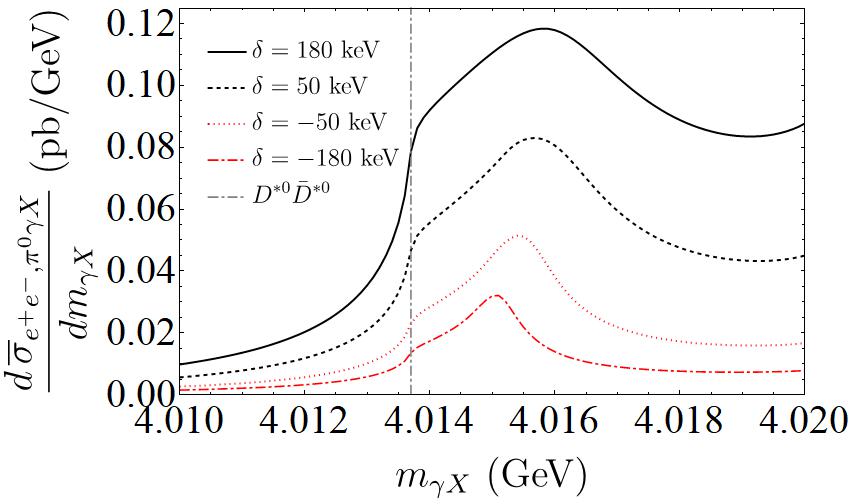}
 \includegraphics[width=8cm]{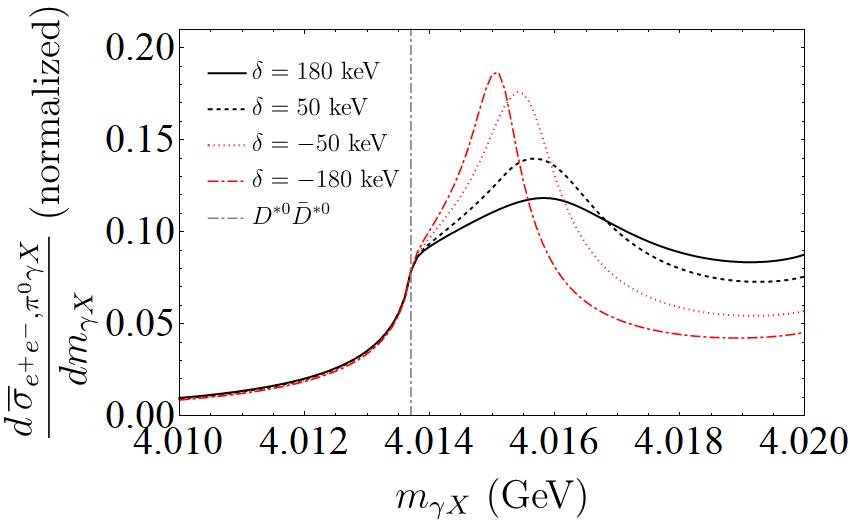}
 \caption{Left:
 the $\gamma\X$ distribution for the $\ee\to\pi^0\gamma\X$ with different $\X$ masses convolved with the Flatt\'e distribution 
 {Eq.~\eqref{eq:flattedist}}.
 The $\ee$ c.m. energy is fixed to be $\sqrt{s}=4.23~\gev$, and the $\X\to \pi^+\pi^- J/\psi$ branching fraction has been taken into account.
 Right:
 the $\ee\to\pi^0\gamma\X$ cross section normalized with the value at $m_{\gamma X}=m_{D^{*0}}+m_{\bar D^{*0}}$ of $\delta=180~\kev$.
 In both panels, the vertical line is the $D^{*0}\bar D^{*0}$ threshold.
  The Flatt\'e parameters with different $\X$ binding energies are given in Table~\ref{tab:fltparameters}.
}
 \label{fig:res1}
\end{figure}

The $\gamma\X$ distribution of the differential $\ee\to\pi^0\gamma\X$ cross section smeared with the Flatt\'e distribution Eq.~\eqref{eq:flattedist} with the parameter sets in Table~\ref{tab:fltparameters} is given in Fig.~\ref{fig:res1}.
{The range of the smearing in Eq.~\eqref{eq:conv1}, twice of the half-maximum width from the peak, is $\tilde{m}_X\in m_{X}\pm 200~\kev$ here.}
The $\gamma\X$ distributions with a few different masses ($\delta$ values) of the $\X$ are shown in the left panel,
and those normalized to the value at $m_{\gamma X}=m_{D^{*0}}+m_{\bar D^{*0}}$ with $\delta=180~\kev$ is also given in the right panel of Fig.~\ref{fig:res1} to make the comparison of the line shapes easier.

The distribution ${d\,\overline{\sigma}_{\ee,\pi^0\gamma X}/dm_{\gamma X}}$, which involves the $\X$ decay into the $\pi^+\pi^-J/\psi$ mode, is the order 0.01~pb/GeV {as in the left panels of Figs.~\ref{fig:res2} and \ref{fig:res1}} within $\delta=\pm 180~\kev$. 
As one can see in the left panel of Fig.~\ref{fig:res1},
the magnitude is bigger with larger $\delta$.
In the right panel of Fig.~\ref{fig:res1}, one can see that the peak of the TS looks more clear with a negative $\delta$ compared with that with a positive $\delta$ or the line in Fig.~\ref{fig:res2}.
The peak positions for the $\delta=-50~\kev$ and $-180~\kev$ cases are $4.0155~\gev$ and $4.015~\gev$, respectively, which are dictated by the TS whose location can be easily obtained using the master formula in Ref.~\cite{Bayar:2016ftu}.
On the other hand, the peak around $m_{\gamma X}=4.016~\gev$ with $\delta>0$ is a remnant of the TS
because the TS is in the complex plane in this case even when the $D^{*0}$ width is neglected.
Thus, the peak is sensitive to the binding energy particularly with $\delta<0$ as can be seen from the figure. As studied in Ref.~\cite{Ortega:2020ayw}, even after considering the energy resolution, the shapes can still be distinguished for different binding energies.

Let us make a comment on the uncertainties of the order of magnitude.
The uncertainty of the $\ee\to\pi^0(D^*\bar D^*)^0$, which is used to fix the parameter $g_0g_1g_2$, is about 10\%,
the uncertainty of the $D^*\to\gamma D$ part is only a few percent referring to the relative errors of the $D^{*+}$ full width and the $D^*$ branching ratios~\cite{Zyla:2020zbs}, and the composition of the $\X$ other than $D\bar D^*$ would give an uncertainty of a few tens of percents to the coupling constant $g_4^2$.
Then, the uncertainties of the cross section are expected be about a few tens of percents.

Integrating the differential cross section in Figs.~\ref{fig:res2} and \ref{fig:res1} over the $m_{\gamma X}$ region between 4.01 and 4.02 GeV, we get 
{$\mathcal{O}(0.1~{\rm fb})$}. 
Such a small cross section implies that measuring the $\gamma \X$ line shape of the $\ee\to\pi^0\gamma\X$ process would be very challenging.

\subsection{\texorpdfstring{$\ppb\to\gamma\X$}{ppbar to gamma X}}

\begin{figure}[t]
 \centering
 \includegraphics[width=8cm]{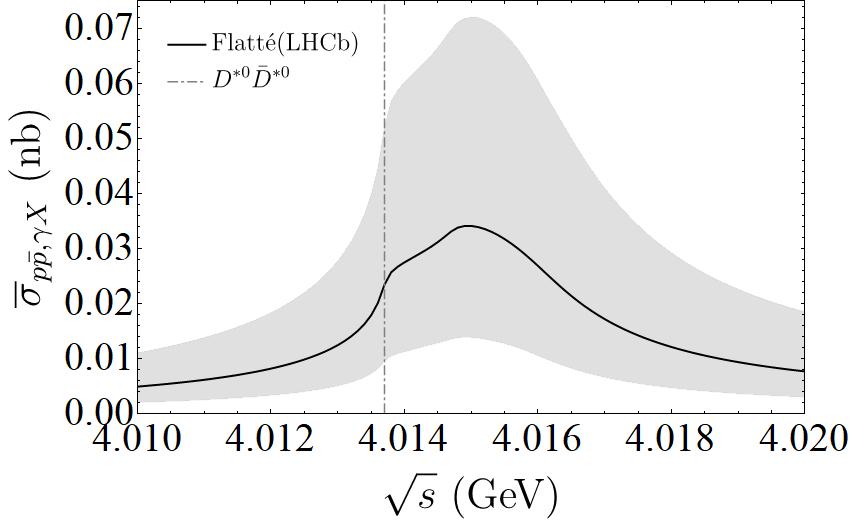}
 \includegraphics[width=8cm]{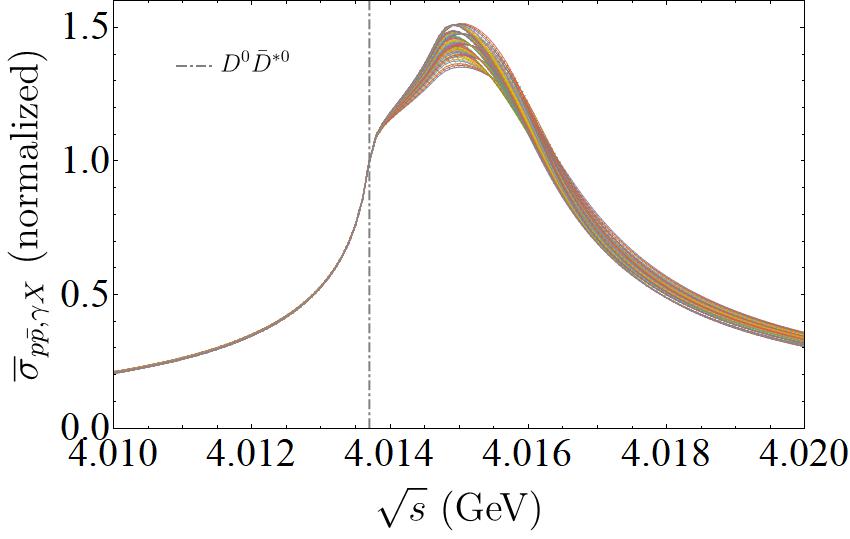}
 \caption{
 Left: the $\gamma\X$ distribution for the $\ppb\to\gamma\X$ smeared with the Flatt\'e distribution.
 The gray error band is given by the parameter errors of the Flatt\'e amplitude of the LHCb analysis~\cite{Aaij:2020qga}.
 {Right: the plot of the $\gamma\X$ distribution in the $\ppb\to\gamma\X$ process with the parameter sets within the Flatt\'e-parameter errors of Ref.~\cite{Aaij:2020qga} normalized with the value at the $D^{*0}\bar D^{*0}$ threshold.}
 In both panels, the $D^{*0}\bar D^{*0}$ threshold is shown with the gray dash-dotted line.
 }
 \label{fig:res4}
\end{figure}

The plot of the $\ppb\to\gamma\X$ cross section convolved with the Flatt\'e distribution in Eq.~\eqref{eq:flattedist}, ${\overline{\sigma}_{\ppb,\gamma X}}$ [note that the $\X\to \pi^+\pi^- J/\psi$ branching fraction has been taken into account as before], as a function of the $\ppb$ c.m. energy, $\sqrt{s}$ is given in Fig.~\ref{fig:res4}.
The Flatt\'e parameters from the LHCb analysis are used~\cite{Aaij:2020qga}.
The distribution is similar to the analogous one for the $\ee\to\pi^0\gamma\X$ in Fig.~\ref{fig:res2}, peaking at $m_{\gamma X}=4.015~\gev${, and the line shape is only marginally changed within errors of the Flatt\'e parameters.}

\begin{figure}[t]
 \centering
 \includegraphics[width=8cm]{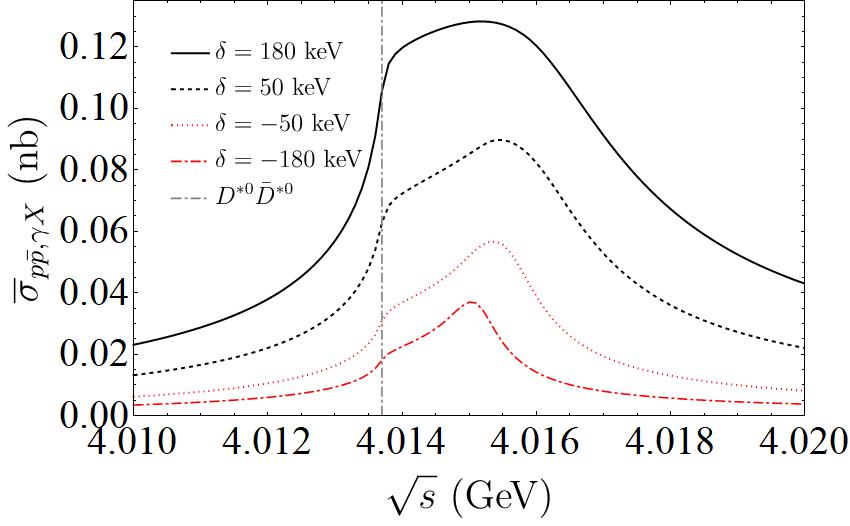}
 \includegraphics[width=8cm]{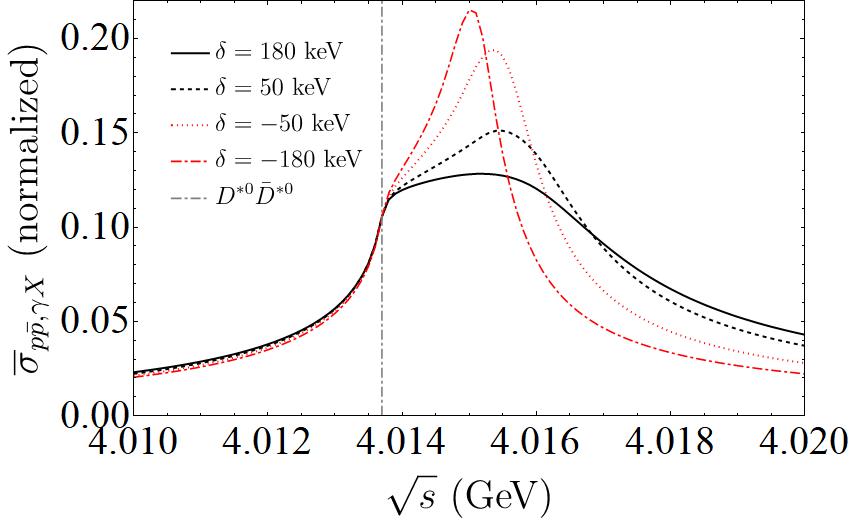}
 \caption{Left:
 the $\gamma\X$ distribution for the $\ppb\to\gamma\X$ with different $\X$ binding energies
 as a function of the $\ppb$ c.m. energy $\sqrt{s}$. The $\X\to \pi^+\pi^- J/\psi$ branching fraction has been taken into account.
 Right:
 the $\ppb\to\gamma\X$ cross section normalized with the value at $m_{\gamma X}=m_{D^{*0}}+m_{\bar D^{*0}}$ of $\delta=180~\kev$.
 In both panels, the vertical line is the $D^{*0}\bar D^{*0}$ threshold. 
 The Flatt\'e parameters with different $\X$ binding energies are given in Table~\ref{tab:fltparameters}.
 }
 \label{fig:res3}
\end{figure}

The plot of the $\ppb\to\gamma\X$ cross section, ${\overline{\sigma}_{\ppb,\gamma X}}$, with the parameter sets in Table~\ref{tab:fltparameters} is given in the left panel of Fig.~\ref{fig:res3}, and the right panel of Fig.~\ref{fig:res3} is the plot with all line shapes normalized to that of $\delta=180$~keV at the $D^{*0}\bar D^{*0}$ threshold as the right panel of Fig.~\ref{fig:res1}.
The $\gamma\X$ invariant mass distribution of the $\ppb\to\gamma\X$ process is qualitatively the same as the $\ee\to\pi^0\gamma\X$ case, since the singularities are the same.
The cross section increases with larger $\delta$, 
and the peak structure looks more significant with $\delta<0$.
The lines of $\delta=-50~\kev$ and $-180~\kev$ show a clear peak structure due to the TS in the physical region:
the peak of $\delta=-50~\kev$ is at $4.0155~\gev$ and that of $\delta=-180~\kev$ is at $4.015~\gev$ as in Fig.~\ref{fig:res1}.
Comparing the distributions of $\delta=50~\kev$ and $\delta=180~\kev$,
the enhancement at $m_{\gamma X}=4.016~\gev$ in the $\gamma\X$ distribution with $\delta=50~\kev$ is more clear since the TS is closer to the physical region.

\begin{figure}[t]
 \centering
 \includegraphics[width=8cm]{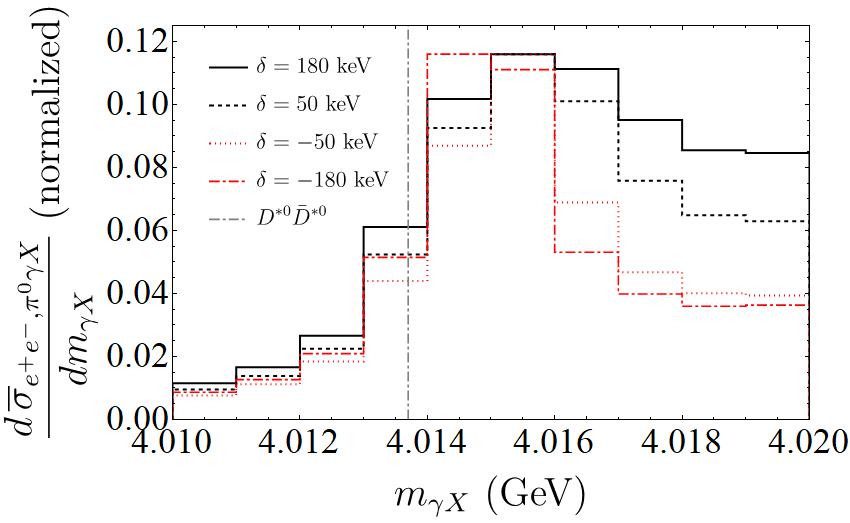}
 \includegraphics[width=8cm]{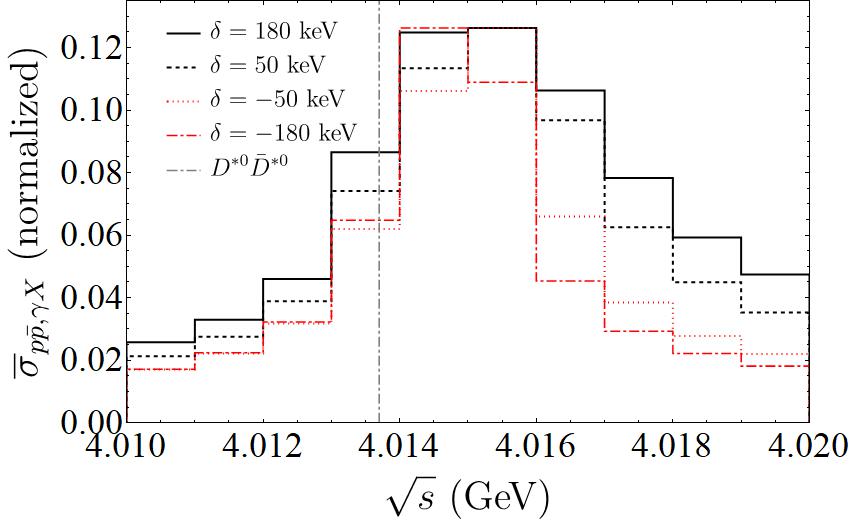}
 \caption{The bin-averaged histogram of $d\,\overline{\sigma}_{\ee,\pi^0\gamma X}/dm_{\gamma X}$ (left) and $\overline{\sigma}_{p\bar p,\gamma X}$ (right).
 The energy bin is $1~\mev$, and the magnitudes are normalized at their maximal values.
 The $D^{*0}\bar D^{*0}$ threshold is shown as the gray dash-dotted line.
 }
 \label{fig:binaverage}
\end{figure}

About the cutoff $\Lambda$ in the form factor Eq.~\eqref{eq:ff} for the $\ppb\to\bar D^*D^*$ transition, 
$\Lambda=2~\gev$ is used in the plot of Figs.~\ref{fig:res4} and \ref{fig:res3}.
Varying the cutoff $\Lambda$ within $\Lambda=2.0\pm 0.2~\gev$, the cross section changes by a factor of 2 compared to the value with $\Lambda=2~\gev$
with the same line shape, indicating a large uncertainty in the estimate of  the $\ppb\to\gamma\X$ cross section in addition to that in the $\X\to D\bar D^*$ coupling $g_4$. 
Nevertheless, the order of magnitude should be reliable, and we expect $\overline{\sigma}_{p\bar p, \gamma X}$ to be of $\mathcal{O}(10~{\rm pb})$ for $\sqrt{s}\sim 4015$~MeV. 
From Ref.~\cite{PANDA:2018zjt}, the integrated luminosity of PANDA at $\sqrt{s}=3872$~MeV is about 2~fb$^{-1}$ in five months. 
Assuming the same integrated luminosity of 2~fb$^{-1}$ in the energy region from 4010 to 4020~MeV, $\mathcal{O}(2\times10^4)$ events are expected to be collected for the $X(3872)$ in the $J/\psi\pi^+\pi^-$ mode. Considering further the reconstruction of the $J/\psi$ from the $e^+e^-$ and $\mu^+\mu^-$ pairs, each of which has about a branching fraction of about 6\%~\cite{Zyla:2020zbs}, we expect that $\mathcal{O}(2\times 10^3)$ events can be collected at PANDA.
According to the Monte Carlo simulation in Ref.~\cite{Guo:2019twa}, a high-precision measurement of the $\X$ binding energy is foreseen even after further smearing due to the energy resolution is taken into account~\cite{Ortega:2020ayw}. In particular, such a smearing effect at PANDA will be very small since the energy resolution can reach the level of 100~keV~\cite{PANDA:2018zjt,Lehrach:2005ji}.\footnote{The beam energy resolutions for the high luminosity and high resolution modes of the High Energy Storage Ring are 167.8~keV and 33.6~keV, respectively~\cite{PANDA:2018zjt,Lehrach:2005ji}.}
To make a better comparison with the forthcoming experimental data, we show in Fig.~\ref{fig:binaverage} the histograms of $d\,\overline{\sigma}_{\ee,\pi^0\gamma X}/dm_{\gamma X}$ and $\overline{\sigma}_{p\bar p,\gamma X}$ averaged over each energy bin.
In the plot, the energy bin size is fixed to be $1~\mev$ and the magnitude of the histogram is normalized at their maximal values.
The histograms with different $\delta$ can be distinguished from the peak position and the shape.
Particularly, the line shape with positive $\delta$ is characterized by a longer tail at larger $\gamma \X$ energies.

\section{Summary}
\label{sec:summary}

In this paper, we have estimated the cross sections for the production of $\gamma\X$ from a short-distance $D^{*0}\bar D^{*0}$ source. A measurement of the $\gamma\X$ line shape was proposed to achieve an unprecedented precision in determining the $\X$ binding energy~\cite{Guo:2019qcn}.
We focused on two processes in this paper: $\ee\to\pi^0\gamma\X$ and $\ppb\to\gamma\X$.
The $\gamma\X$ invariant mass distributions for these two processes were computed, which clearly show a special peak sandwiched between the $D^{*0}\bar D^{*0}$ threshold and  the triangle singularity of the $D^{*0}\bar D^{*0}D^0/\bar D^{*0}D^{*0}\bar D^0$ loops.
The obtained line shapes with different $\X$ binding energies can be distinguished from each other in both the $\ee$ and $\ppb$ processes:
the peak is more narrow when the $\X$ mass is above the $D^0\bar D^{*0}$ threshold.
Convolving the distributions with the spectral function of the $\X$ does not change the conclusion, and the effect of smearing is marginal considering a width of order 100~keV for the $\X$.

In the $\ee\to\pi^0\gamma\X$ reaction, the $\Z$ resonance is introduced, and it is found that this resonance does not essentially change the peak structure caused by the TS.
For the c.m. energy of the $e^+e^-$ pair fixed at 4.23~GeV, with inputs from the BESIII measurements of the $\ee\to\pi^0(D^*\bar D^*)^0$~\cite{Ablikim:2015vvn}, we find that the cross section $\sigma(e^+e^-\to\pi^0\gamma X(3872)) \times \br (X(3872)\to\pi^+\pi^-J/\psi)$ is 
{$\mathcal{O}(0.1~{\rm fb})$} with the $\gamma\X$ invariant mass integrated from 4.01 to 4.02~GeV.
For the $p\bar p\to\gamma X(3872)$, the cross section is much larger. Considering a $\Lambda_c$ exchange to produce $D^{*0}\bar D^{*0}$ from the $p\bar p$ collisions, it is estimated to be $\sigma(p\bar p\to\gamma X(3872))\times \br(X(3872)\to\pi^+\pi^-J/\psi)=\mathcal{O}(10~{\rm pb})$. 
This result indicates that while it is hard to measure $e^+e^-\to\pi^0\gamma X(3872)$, plenty of events can be collected for $p\bar p\to\gamma X(3872)$ at the PANDA experiment.
A precise determination of the $\X$ binding energy is foreseen, which can definitely shed new light into understanding this most mysterious charmoniumlike particle.

\begin{acknowledgments}

This work is supported in part by the National Natural Science Foundation of China (NSFC) under Grants No.~11835015, No.~11947302 and No.~11961141012, by  the NSFC and the Deutsche Forschungsgemeinschaft (DFG) through the funds provided to the Sino-German Collaborative Research Center ``Symmetries and the Emergence of Structure in QCD'' (NSFC Grant No. 11621131001, DFG Grant No. CRC110),
by the Chinese Academy of Sciences (CAS) under Grants No.~XDB34030303 and No.~QYZDB-SSW-SYS013, and
by the CAS Center for Excellence in Particle Physics (CCEPP). 
S.S. is also supported by the 2019 International Postdoctoral Exchange Program, 
and by the CAS President's International Fellowship Initiative (PIFI) under Grant No.~2019PM0108.

\end{acknowledgments}

\appendix
\section{\texorpdfstring{$\ee\to\pi^0\gamma\X$}{eetopi0gamx} and \texorpdfstring{$\ppb\to\gamma\X$}{ppbtogamx} amplitudes}

\label{app:1}
With the $\ee\to\pi^0\gamma\X$ amplitude in Eq.~\eqref{eq:ampeepgx} and the momentum assignment in Fig.~\ref{fig:fig1}, we have
\begin{align}
 \mM_{\ee,\pi^0\gamma X}=&-2e^3g_0g_1g_2g_3g_4D_\gamma^{-1}(s)D_{\psi}^{-1}(s)D_{Z_c}^{-1}(m_{\gamma X}^2)\bar{v}\gamma_{\beta''}u[P_\psi]^{\beta''\beta'}[P_{Z_c}]_{\beta'\beta}\epsilon^{\alpha\beta\gamma\delta}(p_{Z_c})_\alpha\notag\\
 &\times\int\frac{d^4l}{(2\pi)^4}\dtri^{-1} \left[-g_{\gamma\rho}+\frac{(p_{D^{*0}})_{\gamma}(p_{D^{*0}})_{\rho}}{m_{D^{*0}}}\right] \left[-g_{\delta\tau}+\frac{(p_{\bar D^{*0}})_{\delta}(p_{\bar D^{*0}})_{\tau}}{m_{\bar D^{*0}}}\right] \nonumber\\
 &\times
 \epsilon^{\mu\nu\rho\sigma}(p_{D^{*0}})_\mu(p_\gamma)_\nu(\epsilon_\gamma^*)_\sigma(\epsilon_X^*)^\tau\notag\\
 =&-2e^3g_0g_1g_2g_3g_4D_\gamma^{-1}(s)D_{\psi}^{-1}(s)D_{Z_c}^{-1}(m_{\gamma X}^2)\bar{v}\gamma_{\beta''}u[P_\psi]^{\beta''\beta'}[P_{Z_c}]_{\beta'\beta}\epsilon^{\alpha\beta\gamma\delta}(k_1)_\alpha\notag\\
 &\times\int\frac{d^4l}{(2\pi)^4}\dtri^{-1}(-g_{\gamma\rho}) \left(-g_{\beta\tau}+\frac{l_\beta l_\tau}{m_{\bar D^{*0}}^2}\right)
 \epsilon^{\mu\nu\rho\sigma}(k_1+l)_\mu(k_1-k_2)_\nu(\epsilon_\gamma^*)_\sigma(\epsilon_X^*)^\tau.
\end{align}
The $\ppb\to\gamma\X$ amplitude Eq.~\eqref{eq:ppbt1} with the particle momenta assigned as in Fig.~\ref{fig:ppbtogamX} is reduced to
\begin{align}
 \mM_{\ppb,\gamma X}^{(D^{*0}\bar D^{*0}D^0)}
 =&\int\frac{d^4l}{(2\pi)^4}\frac{eg_3g_4F_{p,\bar D^*\Lambda_c}^2}{m_{\Lambda_c}}\bar{v}(ig_v\gamma^{\mu'})(ig_v\gamma^{\mu})u\notag\\
 &\times\dtri^{-1}(-g_{\mu'\gamma}) \left[-g_{\mu\tau}+\frac{(p_{\bar D^{*0}})_{\mu}(p_{\bar D^{*0}})_{\tau}}{m_{\bar D^{*0}}^2}\right]\epsilon^{\alpha\beta\gamma\delta}(p_{D^{*0}})_\alpha(p_\gamma)_\beta(\epsilon_\gamma^*)_\delta(\epsilon_X^*)^\tau\\
 =&-\int\frac{d^4l}{(2\pi)^4}\frac{g_v^2eg_3g_4F_{p,\bar D^*\Lambda_c}^2}{m_{\Lambda_c}}\bar{v}\gamma^{\mu'}\gamma^\mu u\notag\\
 &\times\dtri^{-1}(-g_{\mu'\gamma})(-g_{\mu\tau}+\frac{l_{\mu}l_{\tau}}{m_{\bar D^{*0}}^2})\epsilon^{\alpha\beta\gamma\delta}(k_1+l)_\alpha(k_1-k_2)_\beta(\epsilon_\gamma^*)_\delta(\epsilon_X^*)^\tau,
\end{align}
and the amplitude of the $\bar D^{*0}D^{*0}\bar D^0$ loop, Eq.~\eqref{eq:ppbt2}, gives
\begin{align}
 \mM_{\ppb,\gamma X}^{(\bar D^{*0}D^{*0}\bar D^0)}
 =&-\int\frac{d^4l}{(2\pi)^4}\frac{eg_3g_4F_{p,\bar D^*\Lambda_c}^2}{m_{\Lambda_c}}\bar{v}(ig_v\gamma^{\mu'})(ig_v\gamma^{\mu})u\, \epsilon^{\alpha\beta\gamma\delta}(p_{\bar D^{*0}})_\alpha(p_\gamma)_\beta(\epsilon_\gamma^*)_\delta(\epsilon_X^*)^\tau\notag\\
 &\times\dtri^{-1} \left[-g_{\mu\gamma}+\frac{(p_{\bar D^{*0}})_{\mu}(p_{\bar D^{*0}})_{\gamma}}{m_{D^{*0}}^2}\right] \left[-g_{\mu'\tau}+\frac{(p_{D^{*0}})_{\mu'}(p_{D^{*0}})_{\tau}}{m_{D^{*0}}^2}\right]\notag\\
 =&\int\frac{d^4l}{(2\pi)^4}\frac{g_v^2eg_3g_4F_{p,\bar D^*\Lambda_c}^2}{m_{\Lambda_c}}\bar{v}\gamma^{\mu'}\gamma^\mu u\notag\\
 &\times \dtri^{-1}(-g_{\mu\gamma})\left(-g_{\mu'\tau}+\frac{l_{\mu'}l_{\tau}}{m_{D^{*0}}^2}\right)\epsilon^{\alpha\beta\gamma\delta}(k_1+l)_\alpha(k_1-k_2)_\beta(\epsilon_\gamma^*)_\delta(\epsilon_X^*)^\tau.
\end{align}
Adding these two terms, we get
\begin{align}
 \mM_{\ppb,\gamma X}^{(D^{*0}\bar D^{*0}D^0)}+\mM_{\ppb,\gamma X}^{(\bar D^{*0}D^{*0}\bar D^0)}=&\int\frac{d^4l}{(2\pi)^4}\frac{g_v^2eg_3g_4F_{p,\bar D^*\Lambda_c}^2}{m_{\Lambda_c}}\bar{v}[\gamma^{\mu'},\gamma^\mu] u\dtri^{-1}g_{\mu\gamma} \left(-g_{\mu'\tau}+\frac{l_{\mu'}l_{\tau}}{m_{D^{*0}}^2}\right)\notag\\
 &\times\epsilon^{\alpha\beta\gamma\delta}(k_1+l)_\alpha(k_1-k_2)_\beta(\epsilon_\gamma^*)_\delta(\epsilon_X^*)^\tau.
\end{align}

\section{Pole dominance in the \texorpdfstring{$(\bar D^0D^{*0}+D^0\bar D^{*0})\to J/\psi \pi^+\pi^-$}{DDstar2Jpsirho}}

\label{app:jpsirho}

Here we show that the $(\bar D^0D^{*0}+D^0\bar D^{*0})\to J/\psi \pi^+\pi^-$ process is dominated by the $X(3872)$ pole if the events selection is restricted to a small region around the $X(3872)$ mass, despite that the $X(3872)\to J/\psi\pi^+\pi^-$ receives a suppression from isospin breaking.

\begin{figure}
    \centering
    \includegraphics[width=0.5\textwidth]{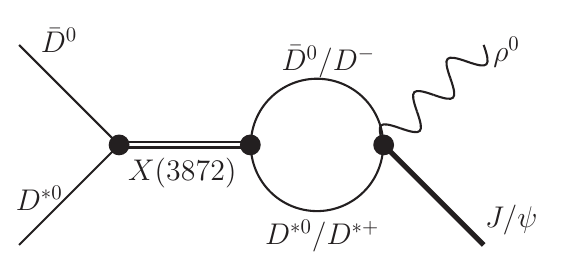}
    \caption{Isospin breaking process of $(\bar D^0D^{*0}+D^0\bar D^{*0})\to X(3872)\to J/\psi \pi^+\pi^-$. The charge conjugated charmed-meson pairs are not shown.}
    \label{fig:jpsirho}
\end{figure}
Since the $J/\psi\pi^+\pi^-$, which can be treated as from the $J/\psi\rho^0$, is an isospin vector, it is not essential in the formation of the $X(3872)$. Therefore, its contribution to the $X(3872)$ can be treated perturbatively.
The leading term in the Laurent expansion of the full amplitude for $(\bar D^0D^{*0}+D^0\bar D^{*0})\to J/\psi \pi^+\pi^-$ contains the $X(3872)$ pole, which couples to $J/\psi \rho^0$ as shown in Fig.~\ref{fig:jpsirho}~\cite{Gamermann:2009uq}.
The isospin breaking comes from the difference between the contributions from the charged and neutral charmed-meson loops in the figure.
The amplitude can be written as
\begin{equation}
    \frac{g_0^X}{E+\delta + i\,\Gamma_X/2}\left[ g_0^X G_0(E) V_0 - g_c^X G_c(E) V_c \right],
\end{equation}
where $E$ is the difference between the $J/\psi \pi^+\pi^-$ invariant mass and the $D^0\bar D^{*0}$ threshold, $g_0^X$ and $g_c^X$ are the effective coupling constants for the $X(3872)$ couplings to the neutral and charged $D\bar D^*+\bar D D^*$ channels, respectively, $V_{0(c)}$ are the tree-level transition amplitudes from the charmed mesons to the $J/\psi \pi^+\pi^-$ without any pole, and $G_{0(c)}(E)$ is the corresponding two-point scalar loop integral, which, evaluated using a Gaussian regulator, reads~\cite{Guo:2017jvc}
\begin{equation}
    G(E) = \frac{\mu}{2\pi} \left( \frac{\Lambda}{\sqrt{2\pi}} + i \sqrt{2 \mu E} \right),
\end{equation}
with $\mu$ the reduced mass in the relevant charmed-meson channel and $\Lambda$ the cutoff in the Gaussian regulator.
We assume that all the isospin breaking happens through the loops, so that $V_0\approx V_c$.
Then, when the $J/\psi \pi^+\pi^-$ invariant mass is in the vicinity of the $X(3872)$ mass, the relative size of the transition rate for $(\bar D^0D^{*0}+D^0\bar D^{*0})\to J/\psi \pi^+\pi^-$ through the $X(3872)$ pole and that without any pole, i.e., given by the isospin conserving $V_0$, can be estimated as
\begin{equation}
    R \equiv \frac1{2a}\int_{-a}^{a}dE \left| \frac{g_0^X}{E+\delta + i\,\Gamma_X/2}\left[ g_0^X G_0(E) - g_c^X G_c(E) \right] \right|^2 .
    \label{eq:ratio}
\end{equation}
Using the central values of the effective couplings computed in Ref.~\cite{Guo:2014hqa} based on the ratio of the decay amplitudes for $X(3872)\to J/\psi\rho^0$ and $X(3872)\to J/\psi\omega$ extracted in Ref.~\cite{Hanhart:2011tn}, $g_0^X=0.35(0.34)$~GeV$^{-1/2}$ and $g_c^X=0.32(0.26)$~GeV$^{-1/2}$ for $\Lambda=0.5 (1.0)$~GeV.
If the $ J/\psi \pi^+\pi^-$ events are selected within $\pm2$~MeV, i.e. $a=2$~MeV, of the $D^0\bar D^{*0}$ threshold, Eq.~\eqref{eq:ratio} leads to $R \approx 89(166)$ with $\delta=0$ and $\Gamma_X=0.1$~MeV using the inputs for $\Lambda=0.5(1.0)$~GeV. The results with $\delta\in[-180,180]$~keV are of the same order.
If the interval for the events selection is further reduced to $\pm1$~MeV, the ratio becomes even larger: $R \approx 175(328)$.

Therefore, we conclude that it is very reasonable to assume that within the vicinity of the $X(3872)$ mass, the $(\bar D^0D^{*0}+D^0\bar D^{*0})\to J/\psi \pi^+\pi^-$ process is dominated by the $X(3872)$ pole.

\bibliography{biblio}

\end{document}